\documentclass{article}

\usepackage{microtype}
\usepackage{graphicx}
\usepackage{subfigure}
\usepackage{booktabs} 
\usepackage{makecell}
\usepackage[font=small,skip=-10pt]{caption}
\usepackage{xcolor}
\usepackage[colorlinks=true]{hyperref}
\usepackage[xindy]{glossaries}
\usepackage{array,multirow,graphicx}

\usepackage{xcolor,pifont}

\newcommand{\xmark}{\ding{55}}%
\newcommand*\colourcheck[1]{%
  \expandafter\newcommand\csname #1check\endcsname{\textcolor{#1}{\ding{52}}}%
}

\newcommand*\colourxmark[2]{%
  \expandafter\newcommand\csname #2check\endcsname{\textcolor{#2}{\ding{55}}}%
}

\definecolor{ao}{rgb}{0.0, 0.5, 0.0}
\colourcheck{green}

\usepackage[normalem]{ulem} 

\usepackage{soul}
\usepackage{hyperref}
\usepackage{glossaries}

\usepackage{amsmath}
\usepackage{amssymb}
\usepackage{mathtools}
\usepackage{amsthm}

\usepackage[tableposition=top]{caption}

\usepackage[capitalize,noabbrev]{cleveref}

\theoremstyle{plain}

\theoremstyle{definition}

\theoremstyle{remark}

\usepackage[textsize=tiny]{todonotes}

\makeglossaries

\newacronym{ml}{ML}{Machine Learning}
\newacronym{rnn}{RNN}{Recurrent Neural Networks}
\newacronym{cnn}{CNN}{Convolutional Neural Networks}
\newacronym{dbs}{DBS}{Deep Brain Stimulation}
\newacronym{rns}{RNS}{ Responsive Neurostimulation}
\newacronym{gru}{GRU}{Gated Reccurent Unit}
\newacronym{lstm}{LSTM}{Long Short-Term Memory}
\newacronym{tuhz}{TUSZ}{Temple University Hospital \gls{eeg} Seizure Corpus }
\newacronym{mse}{MSE}{Mean Squared Error}
\newacronym{eeg}{EEG}{electroencephalography}
\newacronym{bce}{BCE}{Binary Cross Entropy}
\newacronym{gnn}{GNN}{Graph Neural Networks}

\usepackage{microtype}
\usepackage{graphicx}
\usepackage{subfigure}
\usepackage{booktabs} 

\usepackage{hyperref}

\usepackage[accepted]{icml2024}

\usepackage{amsmath}
\usepackage{amssymb}
\usepackage{mathtools}
\usepackage{amsthm}

\usepackage[capitalize,noabbrev]{cleveref}

\usepackage[textsize=tiny]{todonotes}

\icmltitlerunning{REST: Efficient and Accelerated EEG Seizure Analysis through Residual State Updates}

\begin{document}

\twocolumn[

\icmltitle{REST: Efficient and Accelerated EEG Seizure Analysis \\ through Residual State Updates}


\icmlsetsymbol{equal}{*}

\begin{icmlauthorlist}
\icmlauthor{Arshia Afzal}{1,2}
\icmlauthor{Grigorios Chrysos}{3}
\icmlauthor{Volkan Cevher}{equal,2}
\icmlauthor{Mahsa Shoaran}{equal,1}
 
\end{icmlauthorlist}

\icmlaffiliation{2}{LIONS, EPFL, Switzerland}
\icmlaffiliation{1}{INL, EPFL, Switzerland}
\icmlaffiliation{3}{Department of Electrical and Computer Engineering, University of Wisconsin-Madison, USA}

\icmlcorrespondingauthor{Arshia Afzal}{arshia.afzal@epfl.ch}

\icmlkeywords{Machine Learning, ICML}

\vskip 0.3in ]



\printAffiliationsAndNotice{} 

\begin{abstract}
EEG-based seizure detection models face challenges in terms of inference speed and memory efficiency, limiting their real-time implementation in clinical devices. This paper introduces a novel graph-based residual state update mechanism ({\sc Rest})  for real-time EEG signal analysis in applications such as epileptic seizure detection. By leveraging a combination of graph neural networks and recurrent structures, {\sc Rest} efficiently captures both non-Euclidean geometry and temporal dependencies within EEG data. Our model demonstrates high accuracy in both seizure detection and classification tasks. Notably, {\sc Rest} achieves a remarkable 9-fold acceleration in inference speed  compared to state-of-the-art models, while simultaneously demanding substantially less memory than the smallest model employed for this task. These attributes position {\sc Rest} as a promising candidate for real-time implementation in clinical devices, such as Responsive Neurostimulation or seizure alert systems.

\end{abstract}

\section{Introduction}

Brain disorders, including epilepsy, present substantial challenges globally, prompting the need for innovative approaches in diagnosis and treatment. Recurrent seizures, recognized as one of the most prevalent neurological emergencies globally \cite{strein2019prevention}, impact approximately 50 million people worldwide \cite{beghi2019global}.

Detecting changes in the rhythms of brain activity through the monitoring of \gls{eeg} signal allows us to pinpoint the onset zone and time of seizures \cite{gotman1990automatic,siddiqui2020review}, making \gls{eeg}  an invaluable and extensively utilized tool for seizure detection and localization. Traditionally, neurological experts perform these tasks, involving the time-consuming process of manually labeling periods spanning from hours to days for each individual patient \cite{aiseiz,ahmedt2020neural}. Several studies have explored the application of \gls{ml} in seizure analysis, aiming to simplify the handling of large seizure datasets for experts \cite{tang2021self,ahmedt2020neural,covert2019temporal,siddiqui2020review}. These studies predominantly focus on deep models, known for their accuracy and suitability for clinical applications.




Taking inspiration from computer vision \cite{voulodimos2018deep}, many studies have applied different variations of \gls{cnn} for seizure detection, as demonstrated in \citet{saab2020weak}. Various versions of \gls{gnn} effectively capture non-Euclidean geometry in datasets like \gls{eeg} signals, contributing to enhanced seizure detection and classification \cite{li2022graph,tang2021self,ho2023self}. Additionally, to enhance the performance of deep neural networks and accounting for time-series nature of brain rhythms, different variations of \gls{rnn} have been utilized in  seizure analyses  \cite{ahmedt2020neural}.

While these models excel in achieving high accuracy in seizure detection and classification tasks, they often struggle with issues such as complexity, inefficient memory usage, and slow inference speeds. One of the main reasons behind this inefficiency lies in structures such as the gating mechanism found in \gls{rnn} models (e.g., \gls{lstm} \cite{lstm} or the presence of deep convolutional layers in \gls{cnn}s and \gls{gnn}s.

Both inference time and memory storage considerations become critically important in the context of modern seizure treatment devices like \gls{rns} and \gls{dbs} \cite{dbs,rns}. These devices, which have shown promise in suppressing seizure attacks, require a small yet accurate ML model to trigger stimulation commands  for symptom suppression \cite{shoaran201616, shin2022neuraltree}. Furthermore, the model must exhibit low inference time in activating the stimulator to ensure its effectiveness  \cite{dbstim, zhu2021closed}. Unfortunately the aforementioned methods do not have such a low inference.

In this study, we introduce {\sc Rest}, a graph-based residual update mechanism designed to efficiently detect both spatial and temporal information from \gls{eeg}. {\sc Rest} captures spatio-temporal dependencies in \gls{eeg} signals without relying on computationally expensive gating mechanisms commonly found in existing models  \cite{lstm,gru,asif2020seizurenet,tang2021self}. The ability to dynamically capture spatial information over time and update the state accordingly contributes to the high accuracy of {\sc Rest} in localizing and detecting seizures. Notably, {\sc Rest} attains comparable accuracy to state-of-the-art models, while achieving significantly faster processing during inference and substantially reducing  computational and memory overhead \footnote{Visit our web site at \href{https://arshiaafzal.github.io/REST/}{{\color{magenta}https://arshiaafzal.github.io/REST/}}}. Our contributions are as follows:

\begin{itemize}
\item We present a novel graph-based residual update mechanism designed to capture spatio-temporal dependencies in \gls{eeg} signals. 

\item We enhance the model's performance while maintaining its small size and rapid detection and classification speed using binary random masking the state and multiple state updates.

\item Our model delivers predictions with an impressive inference latency of  1.29ms. This unmatched inference speed is achieved with a light memory footprint of 37KB. 

\item Our model is 14$\times$ smaller than the smallest competitive models for seizure detection. Remarkably, our architecture can match the performance of the state-of-the-art deep neural networks with less than 10K parameters.
\end{itemize}

\section{Related Work}

Many studies have attempted to develop  \gls{ml} and deep learning models for seizure detection \cite{siddiqui2020review,o2020neonatal,saab2020weak} and classification of seizure types \cite{ahmedt2020neural,ievsmantas2020convolutional,tang2021self}. Here, we examine existing seizure detection and classification models, assessing their strengths and limitations across three key aspects. Firstly, we explore how these studies capture the spatio-temporal features present in \gls{eeg}. Secondly, we delve into the inference  speed and the impact of varying clip lengths on seizure analysis. Lastly, we study the memory requirements and model size of current  models. 

\textbf{Spatio-Temporal Nature of \gls{eeg} Signals:} As introduced earlier, the nature of \gls{eeg} signals involves both spatial and temporal components, which are pivotal for accurate analysis in epilepsy studies. Notably, some studies, like \citet{asif2020seizurenet}, extract spectral features to represent temporal dependencies,  incorporating them into a \gls{cnn} architecture. In contrast, \citet{saab2020weak} employ a \gls{cnn} model that treats \gls{eeg} signals as multi-channel images,  a methodology that does not align with the time-series structure of \gls{eeg}. Recent advancements involve the utilization of various \gls{rnn}  variations or transformers \cite{atenneed} to effectively capture temporal patterns in alignment with the intricate dynamics of \gls{eeg} signals. 

\gls{rnn}s capture temporal dependencies within time-series data by mapping the input \(x(t)\) into a latent space \(h(t)\) and employ recurrence within that space through linear or non-linear transformations. 
Despite their effectiveness in capturing time-series dependencies, \gls{rnn}s suffer from a significant challenge known as gradient vanishing. This issue occurs during backpropagation, causing gradients to diminish and hindering the effective learning of long-range dependencies in sequential data. To address the vanishing gradient problem \cite{gradvanish}, \gls{rnn} variants like \gls{lstm} \cite{lstm} or \gls{gru} \cite{gru} leverage gating mechanisms, introducing different gates that contribute to creating the next state \(h(t)\) from the current input \(x(t)\) and the previous state \(h(t-1)\). \citet{lstmseiz} used an \gls{lstm} based model for seizure detection.

On the other hand, attention-based models or transformers \cite{atenneed} are more complex than \gls{rnn}s. Rather than constructing an explicit state, they directly use previous inputs to predict the future. However, this approach is more memory-intensive and time-demanding due to the necessity of retaining all prior inputs up to a specified time point and storing weights for each input to construct the attention matrix. \citet{trseiz} employed a transformer-based model for the seizure detection task.

In the context of \gls{eeg} analysis where spatial details are critical at each time point, a common strategy is to utilize a \gls{cnn} or graph convolution network independently across all time points, mapping them into a new feature space. This approach is then complemented by \gls{rnn} to capture temporal dependencies. \citet{ahmedt2020neural} further employ a CNN-LSTM model, effectively addressing both spatial and temporal dependencies in \gls{eeg} data.


\begin{table}[h]
\begin{center}
\caption{\label{table:table1} Comparison of seizure detection and classification methods.  \textbf{A)} Capturing the non-Euclidean geometry of \gls{eeg} signals. \textbf{B)} Capturing the temporal behavior of \gls{eeg} signals. \textbf{C)} Evaluated for both short and long-term seizure detection.  \textbf{D)} Runtime efficient \textbf{E)} Memory efficient.} 
\vspace{0.2cm}

\begin{tabular}{ p{4cm}| p{0.2cm} p{0.2cm} p{0.2cm} p{0.2cm} p{0.2cm} }
 Method & A & B & C & D & E \\
  \Xhline{4\arrayrulewidth}
SeizureNet \cite{asif2020seizurenet}   & \textcolor{red}{\xmark} & \greencheck & \textcolor{red}{\xmark} & \textcolor{red}{\xmark}  & \textcolor{red}{\xmark} \\
\hline
Transformer \cite{transseiz}   & \textcolor{red}{\xmark} & \greencheck & \greencheck  & \textcolor{red}{\xmark}  & \textcolor{red}{\xmark} \\
\hline
EEG-CGS \cite{ho2023self}  & \greencheck & \greencheck & \textcolor{red}{\xmark} & \textcolor{red}{\xmark}  & \textcolor{red}{\xmark} \\
\hline
GGN \cite{li2022graph}  & \greencheck & \greencheck  & \greencheck  & \textcolor{red}{\xmark}  & \textcolor{red}{\xmark} \\
\hline
LSTM \cite{lstm}  & \textcolor{red}{\xmark} & \greencheck & \textcolor{red}{\xmark}  & \textcolor{red}{\xmark}  & \textcolor{red}{\xmark}  \\
\hline
CNN-LSTM [1] \cite{ahmedt2020neural}  & \textcolor{red}{\xmark}  & \greencheck  & \textcolor{red}{\xmark} & \textcolor{red}{\xmark}  & \textcolor{red}{\xmark} \\
\hline
CNN-LSTM [2] \cite{lstmseiz}  & \textcolor{red}{\xmark}  & \greencheck  & \textcolor{red}{\xmark} & \textcolor{red}{\xmark}  & \textcolor{red}{\xmark} \\
\hline
DCRNN \cite{tang2021self} & \greencheck & \greencheck  &  \greencheck & \textcolor{red}{\xmark} & \textcolor{red}{\xmark} \\
\hline
{\sc Rest} (Ours)  & \greencheck & \greencheck & \greencheck & \greencheck & \greencheck \\
 \end{tabular}
\end{center}

\end{table}

\begin{figure*}
\begin{minipage}[t]{2\columnwidth} 
 \includegraphics[width=\textwidth,height=8cm]{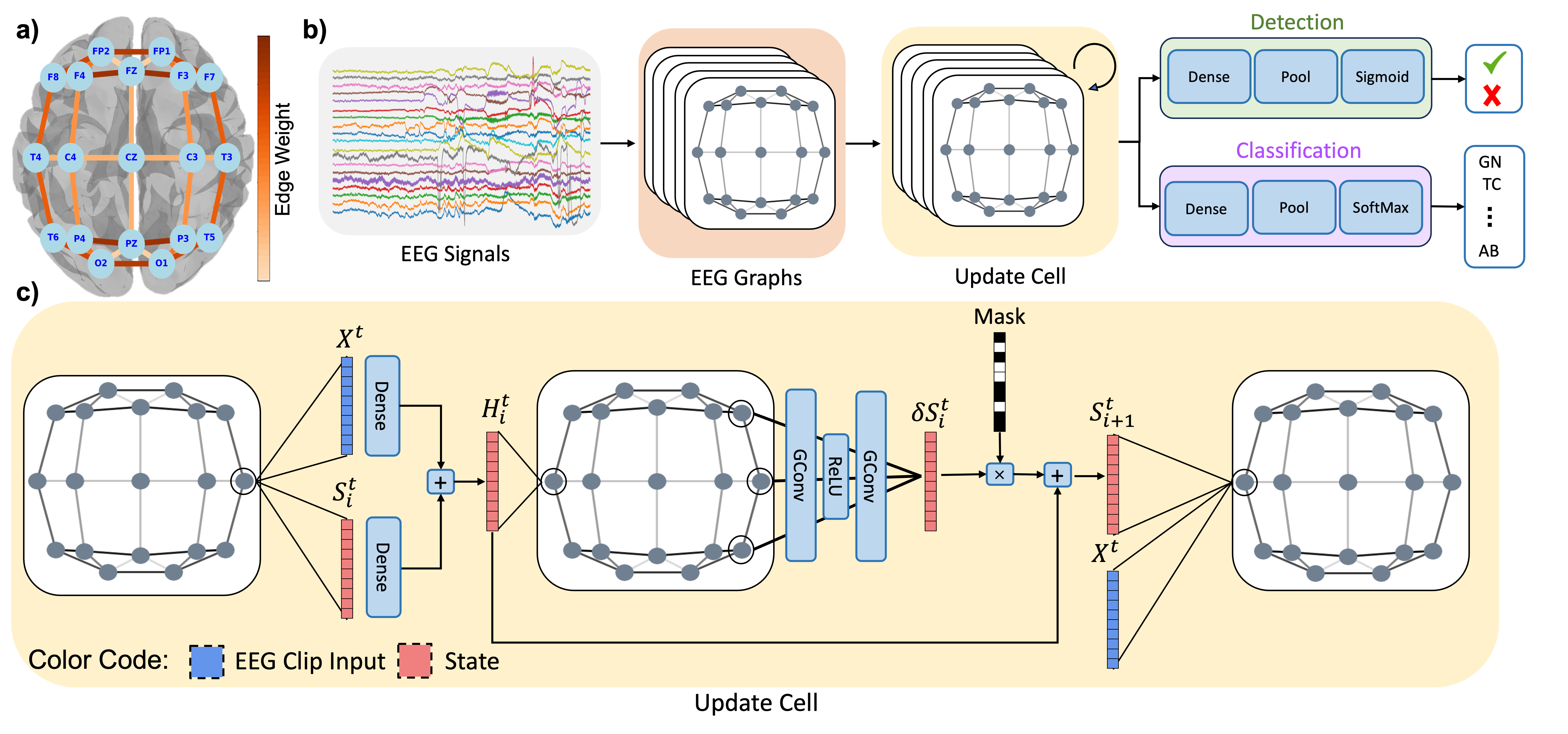}
  \caption{\textbf{(a)} \gls{eeg} electrodes placement based on the 10/20 standard and its constructed distanced based \gls{eeg} graph. Self edges are not shown for better visualization. \textbf{(b)} The {\sc Rest} framework, where raw \gls{eeg} signals undergo preprocessing and are structured as a graph before feeding as input to the model. Following multiple (or single) updates, the model provides the detection or classification result. \textbf{(c)} Single update mechanism of the proposed model. Dense represents the fully connected layer and GConv is the graph convolution. See our web page for more visual results at  \href{https://arshiaafzal.github.io/REST/}{{\color{magenta}https://arshiaafzal.github.io/REST/}}.}
  \label{fig:fig1}
\end{minipage}
\end{figure*}

Nevertheless, these approaches assume Euclidean geometry for \gls{eeg} signals, overlooking the natural geometry of electrode placement (\cref{fig:fig1} a) and brain network connectivity \cite{tang2021self}. Recent studies exploit \gls{gnn}s and graph-based modeling to capture the non-Euclidean geometry of \gls{eeg} signals \cite{tang2021self,ho2023self,covert2019temporal,li2022graph}. For instance, \citet{tang2021self} implement a self-supervised diffusion graph convolution model for both detection and classification tasks. Similarly, \citet{ho2023self} employ a self-supervised graph network for channel anomaly detection. These studies \cite{ho2023self,tang2021self} align more closely with the dynamic changes in \gls{eeg} rhythms by replacing the weights of the \gls{rnn} network with graph convolution filters. This approach represents the evolution of spectral features within each time point of the time-series data, offering a more integrated approach compared to the sequential mapping from \gls{cnn} to \gls{lstm} \cite{ahmedt2020neural}.

\textbf{Significance of Inference Time:} Timely detection of seizure events is essential for the efficacy of closed-loop epileptic treatments such as \gls{rns} and \gls{dbs} \cite{shoaran201616}. To the best of our knowledge, most previous studies either overlook the importance of inference runtime or, as observed in \citet{asif2020seizurenet}, consider a 90ms delay for giving predictions. This delay is still significant, especially for edge devices like RNS and DBS.
 Furthermore, current studies often evaluate models using a limited range of long window sizes, typically exceeding 10 seconds or even 1 minute \cite{tang2021self,saab2020weak}. However, shorter window sizes are preferable for real-time seizure detection and responsive intervention \cite{window, zhu2020resot}. The chosen window size influences a model's ability to localize seizures and its overall detection performance. For instance, a model designed for extended window sizes may lose accuracy in short-term seizure detection scenarios, an aspect that has not been extensively explored in the literature.

\textbf{Memory Requirement in Seizure Detection Models:}
While numerous studies have focused on enhancing the accuracy of seizure detection and classification tasks, the crucial aspect of memory demand remains largely overlooked. For instance, \citet{tang2021self} utilize 240K parameters with complex gating units, \citet{ho2023self} employ 58K for channel anomaly detection, and \citet{asif2020seizurenet} address seizure classification task with a substantial number of 45.94 Million parameters. These examples underscore the need for an efficient model tailored for seizure detection and classification problems, especially one suitable for resource-constrained stimulation devices deployed at the edge, which do not have access to extensive memory storage for model weights and states \cite{zhu2020resot}.

In \cref{table:table1}, we present a summary of current models, highlighting their respective strengths and weaknesses.

\section{Method}
Below, we first formulate the tasks of seizure detection and classification, outlining the graph representation of EEG signals.  Next, we describe the design of {\sc Rest}'s structure using various updating strategies.

\subsection{Seizure Detection and Classification Problem Setting}

Following the preprocessing of raw \gls{eeg} signals and constructing the \gls{eeg} graph, we obtain an \gls{eeg} clip $X$ and a label $y$ for both detection and classification tasks. Here, $X \in \mathbb{R}^{T\times M \times N}$ with $N$ electrodes, $T$ time points and $M$ features per node while $y$ denotes the label. For  detection, the label is binary, whereas for classification, the label falls within the range of \{0,1,2,3,4\} where each class represents a unique seizure type \footnote{The five seizure types include: focal, generalized non-specific, complex partial, absence, and
tonic-clonic.}. The goal for both tasks is to predict the label $y$  based on a given \gls{eeg} clip $X$.

\subsection{EEG Distance Graph Construction}
For each \gls{eeg} clip, we denote a graph as \(\mathcal{G} = \{\mathcal{V} , \mathcal{E} , \mathcal{A} \}\) where \(\mathcal{V} =  \{v_1,...,v_N\} \) represents the nodes corresponding to \gls{eeg} electrodes,  \(\mathcal{E}\) represents the edges, and \(\mathcal{A} \in \mathbb{R}^{N\times N} \) denotes the adjacency matrix of the graph where $N$ is the number of nodes which in case of \gls{eeg} data it is the \gls{eeg} electrodes. 
We build a distance-based \gls{eeg} graph (\cref{fig:fig1}a) that precisely represents the electrode placement geometry in the standard 10/20 system \cite{standard1020}. Unlike  correlation graphs, our graph remains static over time, reducing computations during inference, as the graph structure does not need to be constructed for each input \cite{ho2023self}. Details regarding the choice of $k$ and visualization of distance graphs based on threshold values can be found in \cref{ap:h}.

For a distance graph, the adjacency matrix is constructed using the distance between electrode locations, as in previous studies \cite{tang2021self,li2022graph,ho2023self}. As the \gls{eeg} electrode placements are fixed, the adjacency matrix remains unchanged over time. Thus, for each element \(a_{ij}\in\mathcal{A} \):
\begin{equation}
    a_{ij} = \left\{ 
  \begin{array}{ c l }
    \exp(-\frac{||v_i - v_j||^2}{\sigma^2}) & \quad \textrm{if } ||v_i - v_j|| \leq k, \\
    0                 & \quad \textrm{if }   \textrm{Otherwise} ,
  \end{array}
\right.
\end{equation}
where $\sigma$ is the standard deviation of the distances and $k$ is the Gaussian kernel's threshold \cite{shuman2013emerging}.

\subsection{Residual State Update}
Similar to \gls{rnn}s, {\sc Rest} initially maps the input into a latent space, evolving the state over time to reach the final output. In contrast to \gls{rnn}s, {\sc Rest} updates the state using a novel approach that avoids the complexity of gating mechanisms like \gls{lstm} or \gls{gru}, efficiently addressing the vanishing gradient problem with fewer parameters (details in \cref{ap:b}). For mapping to the state space, {\sc Rest} employs a linear mapping represented as:
\begin{equation}
\label{eq:2}
H^t = WX^t + US^{t-1}.
\end{equation}
Here, \(X^t \in \mathbb{R}^{M \times N} \) represents the input, in our case, the preprocessed \gls{eeg} clip at time point \(t \in [1,...,T]\), and \(S^{t-1} \in \mathbb{R}^{Q \times N} \) is the previous state of the model at time point \(t-1\). \(W \in \mathbb{R}^{Q\times M}\) and \(U\in \mathbb{R}^{Q\times Q}\) are the weights of the affine mapping, with \(Q\) being the state size, while \(H^t \in \mathbb{R}^{Q \times N}\) represents the state of {\sc Rest} prior to the update. Inspired by \citet{resnet}, {\sc Rest} uses a residual mechanism to update its latent state:
\begin{equation}
\label{eq:3}
S^t = H^t + \delta S^t.
\end{equation}
Here, \(S^t\) is the next state of the model and \(\delta S^t\) is the incremental update for the model's state. The critical aspect lies in extracting \(\delta S^t\) to align with the spatial changes in \gls{eeg} dynamics at each time point. For this purpose, we utilize the graph convolution method introduced by \citet{gconv}. We opt for this graph convolution because of its simple structure, which is suited for our application. The graph convolution is defined as follows:
\begin{equation}
\label{eq:4}
{O_{[:,i]}^t} = \sigma \Big(\Theta_{1}H_{[:,i]}^t + \Theta_{2}\sum_{j\neq i} a_{ij}H_{[:,j]}^t \Big),
\end{equation}
where \({O}^t \in \mathbb{R}^{Q \times N}\) is the output of the convolutional filter with \(Q\) features per node. \(\Theta_1 , \Theta_2 \in \mathbb{R}^{Q \times Q}\) parameterize the first and second convolutional filters, \(a_{ij}\) represents the edge (in this case, the adjacency matrix element) between node $i,j \in{1,\ldots,N}$ and \(\sigma\) is the activation function. We denote the graph convolution in \cref{eq:4} as $\mathcal{G}_\Theta(H^t)$. Note that in \cref{eq:4}, the summation is performed over the neighbors of each node. Considering that for non-neighbor nodes, $a_{ij}=0$, we can simplify the sum by taking it over all nodes, implicitly incorporating only the neighbor nodes.

The update for the state, \(\delta S^t\), leveraging the graph convolution, is expressed as follows:

\begin{equation}
\label{eq:5}
\delta S^t = \mathcal{G}_\Theta(H^t).
\end{equation}
This approach aligns well with the spatial dynamics of \gls{eeg} signals. We refer to the process of updating the state of our model using \cref{eq:2,eq:3,eq:5} as the update cell of {\sc Rest} (\cref{fig:fig1} - c).

\subsection{Binary Random Mask: Continuous Dropout during Inference}
To combat overfitting in deep neural networks, Dropout is commonly employed, randomly selecting model parameters during training and retaining all parameters during test-time \cite{dropout}. Drawing inspiration from a similar concept in \citet{nca}, we introduce Binary Masking for state updates, preventing overfitting while enabling the model to learn random state updates. This approach prevent the model to overfit as well as accelerates inference during test-time by skipping computations related to zero-masked feature points in the update.
The state update will simply change as follows:
\begin{equation}
\label{eq:6}
S^{t} = H^{t} + \delta S^{t}\odot B.
\end{equation}
Here, $\odot$ denotes the Hadamard product, and \(B \in \mathbb{R}^{Q \times N}\) is the binary mask with \(B_{ij} \sim \mathcal{B}(p)\) from the Bernoulli distribution, where \(B_{ij}\) takes the value 1 with a probability of \(p\) and can be treated as hyperparameter for the model. 

\begin{table*}[h]
\begin{center}
 \caption{\label{table:table2} Summary of \gls{tuhz} v.2.0.0 Train and Evaluation sets used in this study. Columns represent (from left to right): 1) total number of \gls{eeg} files 2) total Number of patients 3) total number of generalized non-specific (GN) 4)  tonic-clonic (TC) 5) absence (AB) 6) focal (FN), and 7) complex parietal (CP) seizure types in train and evaluation sets.}
\vspace{0.2cm}
\begin{tabular}{ p{2cm}|p{1.8cm}|p{1.8cm}|p{1.5cm}|p{1.5cm}|p{1.5cm}|p{1.5cm}|p{1.5cm} }
&\gls{eeg}-Files & Patients &  \multicolumn{5}{c}{Seizure Type Numbers (Seizure Type Sessions)} \\
&(\% Seizures)&(\% Seizures) & GN  & TC& AB  & FN & CP  \\
\hline
Train & 4664(5.34\%) & 579(36\%) & 335(152) & 30(11)& 50(15)& 1516(496)&279(132) \\
Evaluation & 881(5.82\%) & 43(79\%) & 185(54) & 57(8) & 50(1) & 240(98)&108(32)\\
 \end{tabular}
\end{center}
\end{table*}

\subsection{Multiple Update Mechanism: Escaping the Memory Requirements of Stacked 
\gls{rnn} Layers}

As widely recognized in neural networks, increasing the depth enhances performance by enabling the extraction of more general and complex features \cite{dd}. However, this poses a challenge in \gls{rnn}s, where each additional layer increases memory requirements, not only for storing extra weights but also for additional gates and states.

In our study, we tackle this challenge by modifying {\sc Rest} to employ identical weights for state updates, thus facilitating multiple state updates. Although the graph convolution layer appears repetitive, the effect of binary random mask allows {\sc Rest} to learn to update a new part of the state during each iteration. This adaptation allows {\sc Rest} to align itself with the nature of these random updates, contributing to increased performance and enhanced stability  without affecting  memory requirements.

Thus,  the \cref{eq:2,eq:5,eq:6} will be modified as follows:
\begin{equation}
\label{eq:7}
H^t_i = WX^t + US^{t}_i ,
\end{equation}
\begin{equation}
\label{eq:8}
    S^{t}_{i+1} = H^{t}_i + \delta S^{t}_{i}\odot B.
\end{equation}
Here, the index $i$ denotes the current iteration during which the model updates its state, and $\delta S^{t}_i = \mathcal{G}_\Theta(H^t_i)$. It is crucial to emphasize $X^t$ as the feature input at time point $t$ to prevent the model from diverging into a state and neglecting the input during multiple updates (additional details are provided in the \cref{ap:g}). To update the state for the next time point, the final state obtained after multiple updates becomes the initial state. For instance, after updating the model's state \(I\) times at time point \(t\), the initial state for the next time point \(t+1\) is set as the final state after the last update at time point \(t\) (\(S^{t+1}_0\) = \(S^t_I\)). This enables the model to effectively capture the temporal dynamics across different time points. The proposed framework for the update cell is illustrated in (Figure \ref{fig:fig1}c).

Moreover, previous studies \cite{nca,dynca} have demonstrated that recurrently updating the state of neural networks, similar to {\sc Rest} in structure, for image and texture generation contributes to improved stability. We hypothesize that a similar enhancement can be achieved for seizure detection and classification.

\begin{table}[h]
\begin{center}
 \caption{\label{table:tablechb}  Summary of CHB-MIT Train and Evaluation sets used in this study.}
\begin{tabular}{ p{2cm} | p{1.1cm} | p{1.1cm} | p{2.5cm} }
 & Patients & Seizures  & Recording (hours)  \\
  \Xhline{4\arrayrulewidth}
 Train & 18 & 154 & 732  \\
\hline
Evaluation & 3 & 19  & 91\\
\hline
Test & 3 & 19  & 92.5 \\
 \end{tabular}
\end{center}
\vspace{-0.4cm}
\end{table}

\section{{\sc Rest} \& RNNs}
To better understand the memory efficiency and speed advantages of {\sc Rest} during inference, we compare {\sc Rest} with traditional \gls{rnn}s. As mentioned in Related Work, \gls{rnn}s map the input \(x(t)\) to a hidden state \(h(t)\) and update this state over time using the previous state \(h(t-1)\) and the current input \(x(t)\). We highlight the efficiency and connections between {\sc Rest} and other types of RNNs through the following comparisons:

\textbf{Single Update {\sc Rest} vs. Single-Layer RNN:} First we consider a single \gls{gru} as a representative of \gls{rnn} models, which leverages gating mechanisms to mitigate gradient vanishing. 
For a simple \gls{gru} update, we have the following set of equations:
\begin{equation}
\label{eq:9}
r(t) = \sigma(W_r \cdot [h(t-1), x(t)]),
\end{equation}
\begin{equation}
\label{eq:10}
    z(t) = \sigma(W_z \cdot [h(t-1), x(t)]) ,
\end{equation}
\begin{equation}
\label{eq:11}
    \tilde{h}(t) = \tanh(W_h \cdot [r(t) \odot h(t-1), x(t)]),
\end{equation}  
\begin{equation}
\label{eq:12}
    h(t) = (1 - z(t)) \odot h(t-1) + z(t) \odot \tilde{h}(t) .
\end{equation}
Here,
\( h(t) \) is the hidden state at time \( t \),
\( x(t) \) is the input at time \( t \),
\( \sigma \) is the sigmoid activation function,
\( \odot \) denotes element-wise multiplication,
\( [a, b] \) denotes the concatenation of vectors \( a \) and \( b \), and
\( W_r, W_z, W_h \) represent  the weight matrices.

These equations describe how the hidden state \(h(t)\) is updated over time based on the input and the preceding state. Unlike {\sc Rest}, \gls{gru} relies on three different gates (\(z(t), r(t), \tilde{h}(t)\)) for each state update, requiring twice as much memory as {\sc Rest}, in addition to the storage required for the weights utilized in generating these gates.

Despite \gls{gru}'s memory demands, it not only needs to compute the next state (\(h(t)\)), but also three additional gates (\(z(t), r(t), \tilde{h}(t)\)) as the next state depends on these gates. In contrast, {\sc Rest} relies solely on the update result (\(\delta S^{t}\)), enabling it to rapidly derive the next state by adding it to the previous state, without the need for additional gates.

\textbf{Multi Random Update {\sc Rest} vs. Multi-Layer RNN:}

The remarkable efficiency of {\sc Rest} becomes particularly evident when comparing it with multi-layer \gls{rnn}. In the context of multi-layer \gls{gru}, reaching the final state involves computing a set of equations (\cref{eq:9,eq:10,eq:11,eq:12}) for each layer. This process introduces three times more latency per layer, as each layer has three gates that must be computed to obtain the next state. Furthermore, it requires additional memory to store the hidden state of each layer, especially since it is required for updating the final hidden state of the last layer.

Contrastingly, {\sc Rest} distinguishes itself by reusing the same set of weights for the update cell and state evolution. This  eliminates the need to store the previous state, as  it evolves a distinct state over iterations. Consequently, {\sc Rest} maintains the same memory requirements as a single update, while delivering more accurate results (as discussed in the next section).
It is worth mentioning  that in the context of \gls{eeg} data, all fully connected layers will be replaced by graph convolutions for both {\sc Rest} and \gls{gru}. For example, the combination of \gls{gru} with diffusion graph convolution for a traffic forecasting problem was undertaken by \citet{li2017graph}.

\textbf{Connection of {\sc Rest} Update Cell to Gating Mechanism:}

As shown in \cref{eq:12}, the state update of RNNs, such as GRU, can be expressed as:
\begin{equation}
\label{eq:12new}
    h(t) = h(t-1) + z(t) \odot \left( \tilde{h}(t) - h(t-1) \right).
\end{equation}
This update shares similarities with the {\sc Rest} cell update in \cref{eq:6}. Instead of learning both \(\tilde{h}(t)\) and \(h(t)\) separately, the {\sc Rest} update directly learns \(\tilde{h}(t) - h(t-1)\) as the residual update \(\delta S^{t}\). Additionally, the update gate vector \(z(t)\) is replaced with binary random masking. This substitution reduces the computational and memory overhead required for building \(z(t)\) from the input \(x(t)\) and hidden state \(h(t)\).

\begin{table*}[h]
\begin{center}
\caption{\label{table:table3} Summary of models for seizure detection on the \gls{tuhz} dataset. AUROC of different models is represented along with their memory demands and inference times. }
\begin{tabular}{ p{1.9cm} | p{1cm} p{1cm} p{1cm} p{1.2cm} p{1cm} p{1.4cm} | p{1cm} p{1.1cm} p{1.5cm} }

\multicolumn{1}{l}{} &
\multicolumn{6}{c}{Seizure Detection AUROC (\%)} & \multicolumn{3}{c}{Model Efficiency} \\
  \multicolumn{1}{l|}{Model} & \multicolumn{1}{l}{4-s} & \multicolumn{1}{l}{6-s} & \multicolumn{1}{l}{8-s} & \multicolumn{1}{l}{10-s} & \multicolumn{1}{l}{12-s} & \multicolumn{1}{l|}{14-s} & Size(MB) & \#Param & Inference(ms)\\
 \Xhline{4\arrayrulewidth}
 LSTM & $75.5_{\pm0.3}$ & $76.1_{\pm0.07}$ & $80.1_{\pm0.3}$ & $70.43_{\pm0.02}$ & $77.9_{\pm0.06}$ & $74.24_{\pm0.2}$ & 2.147 & 536K & 3.254\\
\cline{1-10}
GRU & $76.1_{\pm0.02}$ & $78.8_{\pm0.03}$ & $73.2_{\pm0.04}$ & $73.5_{\pm0.02}$  & $80.1{\pm0.1}$ & $77.9_{\pm0.04}$ & 1.61 & 402K & 2.12 \\
\cline{1-10}
 ResNet-LSTM  & $79.1_{\pm0.05}$ & $80.1_{\pm0.2}$ & $75.6_{\pm0.07}$ & $74.3_{\pm0.04}$& $78.8_{\pm0.1}$&$80.0_{\pm0.08}$ & 27.6 & 6.9M& 6.78 \\
\cline{1-10}
 ResNet-Dilation-LSTM & $80.2_{\pm0.08}$ & $76.5_{\pm0.12}$ & $75.9_{\pm0.06}$ & $73.6_{\pm0.03}$ & $77.4_{\pm0.15}$ & $78.2_{\pm0.07}$ & 27.6 & 6.9M& 6.78 \\
\cline{1-10}
 CNN-LSTM & $81.3_{\pm0.1}$  & $78.5_{\pm 0.05}$  & $76.4_{\pm 0.01}$ & $75.4_{\pm 0.05}$ & $75.05_{\pm 0.1}$ & $74.0_{\pm 0.03}$ &  22.8 & 6M & 5.624\\
\cline{1-10}
 DCRNN & $79.7_{\pm0.01}$ & $82.1_{\pm0.04}$ & $80.1_{\pm0.04}$ & $80.0_{\pm0.06}$ & $82.5_{\pm0.1}$ &  $80.12_{\pm0.04}$ & 0.884 & 126K & 9.670\\
\cline{1-10}
 DCRNN w/SS & $\textbf{83.0}_{\pm\textbf{0.08}}$ & $81.8_{\pm0.05}$ & $\textbf{82.7}_{\pm\textbf{0.1}}$ & $82.1_{\pm0.03}$ & ${85.6}_{\pm{0.2}}$ &  ${84.0}_{\pm {0.01}}$ & 1.319 & 330K & 23.25\\
\cline{1-10}
 Transformer & $83.0_{\pm0.02}$ & $82.1_{\pm0.03}$& $82.2_{\pm0.04}$& $\textbf{85.5}_{\pm\textbf{0.07}}$ & $\textbf{86.0}_{\pm\textbf{0.03}}$ & $\textbf{85.1}_{\pm\textbf{0.02}}$ & 0.80 & 120.3K & 2.5 \\
\cline{1-10}
 {\sc Rest}\textsubscript{(DS)} & $75.3_{\pm0.2}$ & $67.0_{\pm0.03}$ & $72.2_{\pm0.07}$ &  $74.1_{\pm0.1}$ & $70.6_{\pm0.04}$ & $70.0_{\pm0.04}$ & \textbf{0.037} & \textbf{8.4K} & \textbf{0.615}\\
 {\sc Rest}\textsubscript{(RS)} & $79.4_{\pm0.03}$ & $81.1_{\pm0.01}$ & $81.0_{\pm0.08}$   & $81.8_{\pm0.02}$ & $80.1_{\pm0.1}$ & $78.1_{\pm0.4}$ & \textbf{0.037} & \textbf{8.4K} & \textbf{0.710}\\
 {\sc Rest}\textsubscript{(RM)} & $82.4_{\pm0.04}$ & $\textbf{82.2 }_{\pm\textbf{0.05}}$& $\textbf{82.7}_{\pm\textbf{0.1}}$ & $83.6_{\pm{0.2}}$ & $83.4_{\pm0.09}$ & $82.0_{\pm0.1}$ & \textbf{0.037} & \textbf{8.4K} & \textbf{1.292} \\
 \end{tabular}
\end{center}
\end{table*}

\begin{table*}[h]
\begin{center}
\caption{\label{table:table3chb} Summary of models for seizure detection on the CHB-MIT dataset. AUROC of different models is represented along with their memory demands and inference times. }
\begin{tabular}{ p{2.3cm} | p{1.1cm} p{1.1cm} p{1.1cm} p{1.2cm} p{1.3cm} | p{1.1cm} p{1.3cm} p{1.5cm} }
 \multicolumn{1}{l}{} &
\multicolumn{5}{c}{Seizure Detection AUROC (\%)} & \multicolumn{3}{c}{Model Efficiency} \\
  \multicolumn{1}{l|}{Model} & \multicolumn{1}{l}{4-s} & \multicolumn{1}{l}{6-s} & \multicolumn{1}{l}{8-s} & \multicolumn{1}{l}{10-s} & \multicolumn{1}{l}{12-s} & Size(MB) & \#Param & Inference(ms) \\
 \Xhline{4\arrayrulewidth} 
 LSTM & $85.5_{\pm0.2}$ & $84.1_{\pm0.4}$ & $81.0_{\pm0.2}$ & $75.2_{\pm0.03}$ & $73.5_{\pm0.08}$ &  2.691 & 627K &  3.56 \\
\cline{1-9}
GRU & $76.1_{\pm 0.3}$ & $78.8_{\pm 0.03}$ &$73.2_{\pm 0.4}$ & $73.5_{\pm 0.01}$ & $80.1_{\pm 0.2}$ &   1.92 & 553K & 2.42  \\
\cline{1-9}
ResNet-LSTM & $77.6_{ \pm 0.2}$ & $82.1_{ \pm 0.14}$ & $79.9_ {\pm 0.3}$ & $76.8_ {\pm 0.4}$ & $81.4_{ \pm 0.17}$   & 29.1 & 7.2M & $6.84 $ \\ 
\cline{1-9}
 ResNet-Dilation-LSTM & $78.2 _{\pm 0.03}$ & $79.8_{ \pm 0.1}$ & $82.3_{ \pm 0.4}$ & $77.6 _{\pm 0.4}$ & $81.2_{ \pm 0.1}$ &  29.1 & 7.2M & 6.84 \\ 
\cline{1-9}
 CNN-LSTM & $86.2_{\pm 0.4}$  & $84.9_{\pm0.2}$ & $80.4_{\pm0.04}$ & $80.35_{\pm0.06}$  &  $77.6_{\pm0.3}$ &   7.6M & 30.23 & 6.432 \\
\cline{1-9}
 DCRNN &  $88.7_{\pm 0.3}$ & $80.0_{\pm0.02}$ & $86.8_{\pm0.06}$ & $88.8_{\pm0.3}$ & $86.5_{\pm0.3}$&  0.591 & 147K & 9.80\\
\cline{1-9}
 Transformer & $80.1_{\pm0.2}$ &$82.3_{\pm0.6}$  & $82.2_{\pm0.04}$ & $85.5_{\pm0.01}$  & $86_{\pm0.17}$   &  0.25 & 52.4K & 6.00 \\
\cline{1-9}
 {\sc Rest}\textsubscript{(DS)} & $89.1_{\pm0.2}$ & $88.5_{\pm0.08}$  & $90.1_{\pm0.1}$  &  $86.3_{\pm0.03}$ & $87.8_{\pm0.5}$  &  \textbf{0.037} & \textbf{9.3K} &   \textbf{1.314}\\
 {\sc Rest}\textsubscript{(RS)} & $92.3_{\pm0.1}$ & $88.7_{\pm0.06}$ &$\textbf{92.1}_{\pm\textbf{0.03}}$ &$\textbf{93.5}_{\pm\textbf{0.02}}$ &   $91.5_{\pm0.02}$& \textbf{0.037} & \textbf{9.3K} &  \textbf{1.314} \\
 {\sc Rest}\textsubscript{(RM)}  & $\textbf{96.7}_{\pm\textbf{0.2}}$  & $\textbf{92.3}_{\pm\textbf{0.04}}$ & $91.4_{\pm0.1}$ & $89.2_{\pm0.4}$ & $\textbf{91.6}_{\pm\textbf{0.03}}$ &  \textbf{0.037} & \textbf{9.3K} & \textbf{1.314} \\
 
 \end{tabular}
\end{center}
\end{table*}

\section{Empirical Results}
\subsection{Setup}

\textbf{Dataset}:
We used two extensive publicly available datasets for the seizure detection and classification task: the Temple University Hospital \gls{eeg} Seizure Corpus (TUSZ) \cite{tuh1,tuh2} and the Children’s Hospital Boston \cite{chb-dataset} dataset. Below is a detailed description of each dataset:

\textbf{\gls{tuhz}}  
This dataset includes a total of 5545 EEG files for training and evaluation. These files encompass five different seizure types. We incorporated all 19 channels for all patients in the standard 10-20 system (Figure \ref{fig:fig1}a).

\textbf{CHB-MIT } 
This dataset comprises recordings from 24 patients, with each patient having data from 9 to 42 sessions, recorded at a sampling rate of 256Hz. The dataset contains a total of 192 seizures. For our study, we included all 19 channels in the standard 10-20 system for the majority of patients, and excluded sessions that had fewer or a higher number of channels.

\textbf{Preprocessing:} In line with  previous studies \cite{tang2021self,saab2020weak}, we resample the \gls{eeg} signals from \gls{tuhz} dataset into 200Hz (256Hz for CHB-MIT dataset) to have consistent sampling frequency among different \gls{eeg}s. Then, we extract non-overlapping window sizes with length $T$ leading to an \gls{eeg} clip $X \in \mathbb{R}^{T\times L \times N}$ with $N=19$ nodes, $L=200$ ($L=256$ for CHB-MIT dataset) features per node, and $T$ time points. After applying the fast Fourier transform on the second dimension of the \gls{eeg} clip and choosing the log amplitude of non-negative frequency components, the final \gls{eeg} clip fused as the input to the models is $X \in \mathbb{R}^{T\times M \times N}$ where $M=100$ ($M=128$ for CHB-MIT dataset). Finally, the features for each node and time point are z-normalized using the mean and variance calculated from 100 (128 for CHB-MIT dataset) feature points along its axis. We examine the presence of a seizure  within an \gls{eeg} clip in the detection task. For classification, we start analyzing each clip 2 seconds before the seizure begins and evaluate the outcomes within a clip duration of $T=10$ seconds. This approach aligns with the annotations of seizure onset, as demonstrated in previous works \cite{ahmedt2020neural,tang2021self}.

 We evaluate models' ability to perform detection tasks across a range of window sizes, spanning from \{4,6,8,10,12,14\} seconds for \gls{tuhz} and \{4,6,8,10,12\} seconds for CHB-MIT. This allows us to evaluate their performance in both short and long-term detection scenarios. For seizure detection task, we used both the seizure and background data, while for the classification task, only the seizure data were used (details in \cref{ap:a}). 

\textbf{Train-Evaluation Split:} The original \gls{tuhz} Train-set was randomly split into training and validation sets with a ratio of 90/10. The \gls{tuhz} eval set served as a standardized evaluation set, consistent with previous studies  \citet{tang2021self}. Further details regarding the data split are provided in Table \ref{table:table2}. For the CHB-MIT dataset, since predefined splits for training, evaluation, and testing are not provided, we randomly selected 80\% of the data for training, 10\% for evaluation, and 10\% for testing. We ensured that patients in each set are unique, preventing the model from being tested on patients included in the training set (details at \cref{table:tablechb}).

\textbf{Baselines:}
To evaluate performance and runtime, we implemented three key baselines widely used in seizure analysis: \textbf{DCRNN} \cite{tang2021self}, with two versions of the model, with and without self-supervision; \textbf{CNN-LSTM} \cite{ahmedt2020neural}; \textbf{LSTM} \cite{lstm}; \textbf{Transformer} \cite{atenneed}; \textbf{GRU} \cite{gru}; and two versions of the \textbf{ResNet-LSTM} model as described in \citet{lee2022real}.

\begin{figure}[h]
\begin{minipage}[t]{1\columnwidth} 
 \includegraphics[width=\textwidth]{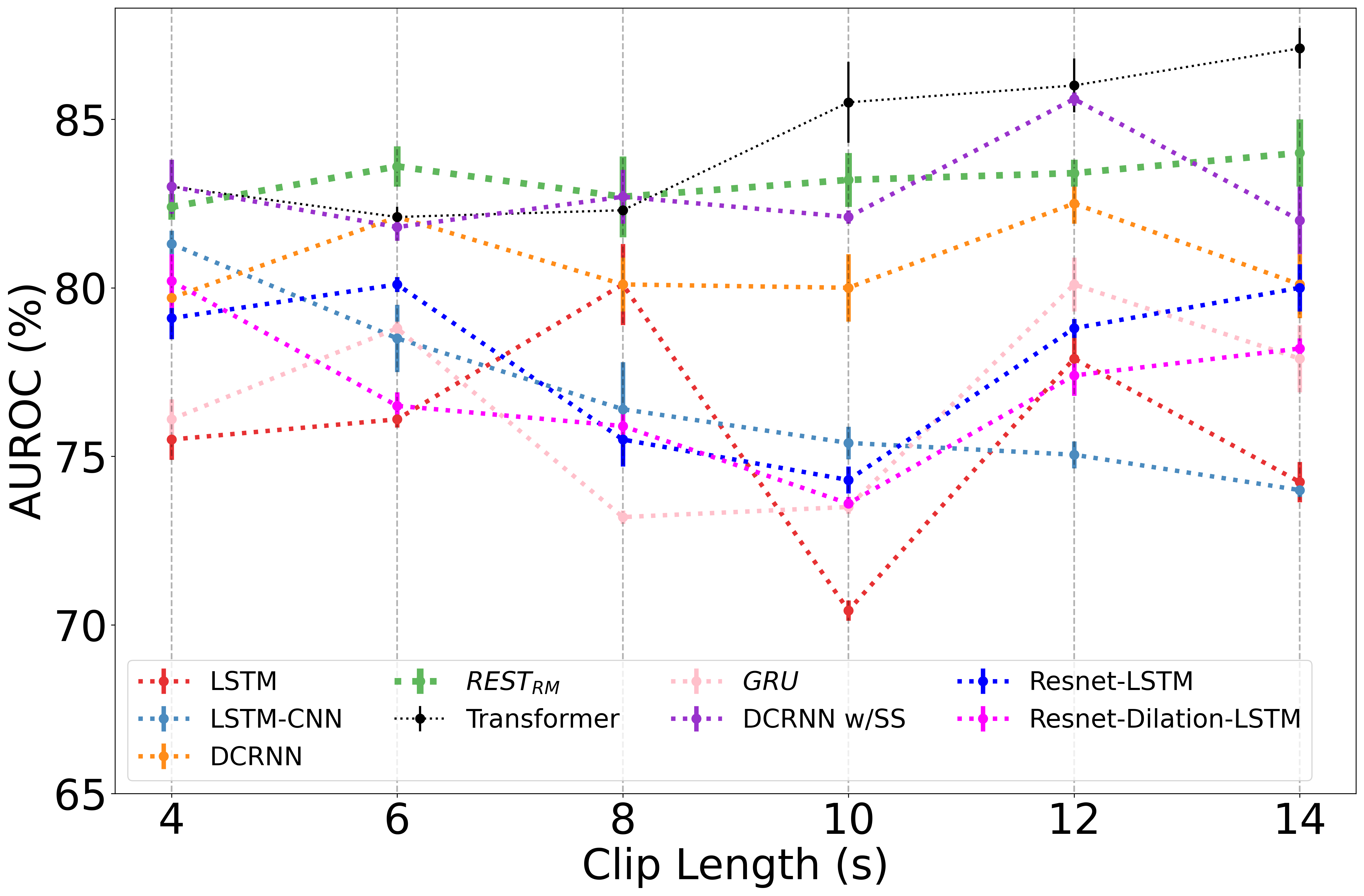}
  \caption{
AUROC comparison among various models for seizure detection across different clip lengths on TUSZ dataset. A flatter line indicates more consistent performance, with error bars representing variation across five random seeds. Higher values on the y-axis correspond to increased accuracy. {\sc Rest}\textsubscript{(RM)} is shown as bold green line to emphasise its stability.}
  \label{fig:fig2}
  \end{minipage}
\end{figure}

\textbf{{\sc Rest} architecture and training:} 
 {\sc Rest} was designed with two graph convolution layers for state updates, the first employing ReLU activation and the second utilizing a linear activation function (Figure \ref{fig:fig1}c).
We evaluate various versions of {\sc Rest}: a) \emph{{\sc Rest}\textsubscript{(DS)}} with a single deterministic update without any masking, b)  \emph{{\sc Rest}\textsubscript{(RS)}} with a single random update (utilizing binary random masking), and c)  \emph{{\sc Rest}\textsubscript{(RM)}} with multiple random updates. 

In the seizure detection task, both Binary Cross Entropy and \gls{mse} loss were employed, with \gls{mse} outperforming Binary Cross Entropy. This result stems from the observation that Binary Cross Entropy prevents residual updates from approaching zero (more details on \cref{ap:e}). For seizure classification, the Cross-Entropy loss was utilized.

\begin{table}[h]
\begin{center}
 \caption{\label{table:table4} Classification Performance, model size and parameter count for different models under the clip length of 10-s.}
 \vspace{0.2cm}
\begin{tabular}{ p{2cm} p{1.5cm} p{1.5cm} p{1.5cm}}
 Model& F1-Score  & Size(MB) & Parameter(\#)  \\
  \Xhline{4\arrayrulewidth}
 LSTM &0.39 & 2.021 & 512K \\
\hline
GRU & 0.44 & 1.92 & 553K\\
\hline
ResNet-LSTM & 0.58 & 30.3 & 7.5M\\
\hline
ResNet-LSTM-Dilation & 0.50 & 30.3 & 7.5M\\
\hline
CNN-LSTM & 0.47 & 23.9 & 6M \\
\hline
DCRNN & 0.54 & 0.506 & 126K\\
\hline
DCRNN w/SS &\textbf{0.62} &  1.40 & 332K  \\
\hline
Transformer & 0.54 & 0.25 & 53K\\
\hline
{\sc Rest}\textsubscript{(DS)} & 0.51 &\textbf{ 0.034 }& \textbf{8.6K}\\
{\sc Rest}\textsubscript{(RS)} &  0.57 &\textbf{ 0.034} & \textbf{8.6K}\\
{\sc Rest}\textsubscript{(RM)} &  0.60 &\textbf{ 0.034} & \textbf{8.6K} \\
 \end{tabular}
\end{center}
\end{table}

We trained all models with 5 different random seeds and averaged the performance on evaluation set over different runs. We utilized ADAM \cite{adam} to optimize the models' parameters, conducting training on a single NVIDIA A100 GPU with a batch size of 128 EEG clips. Training times  for all models across various clip lengths can be found in the \cref{ap:f}.

\textbf{Runtime Comparison:}
To ensure a fair comparison between different models, we adopted the following approach for each model: 
We selected the optimal set of hyperparameters for each clip length based on performance on the validation set.
Here, inference time refers to the time required for each model to provide predictions for one sample of the test data, where each sample is an EEG clip with length $T \in \{4, 6, 8, 10, 12, 14\}$.
We also attempted to shrink the baselines while maintain the same accuracy for both tasks and the details are reported in \cref{ap:i}.

\begin{figure}[h]
\begin{minipage}[t]{1\columnwidth} 
 \includegraphics[width=\textwidth ]{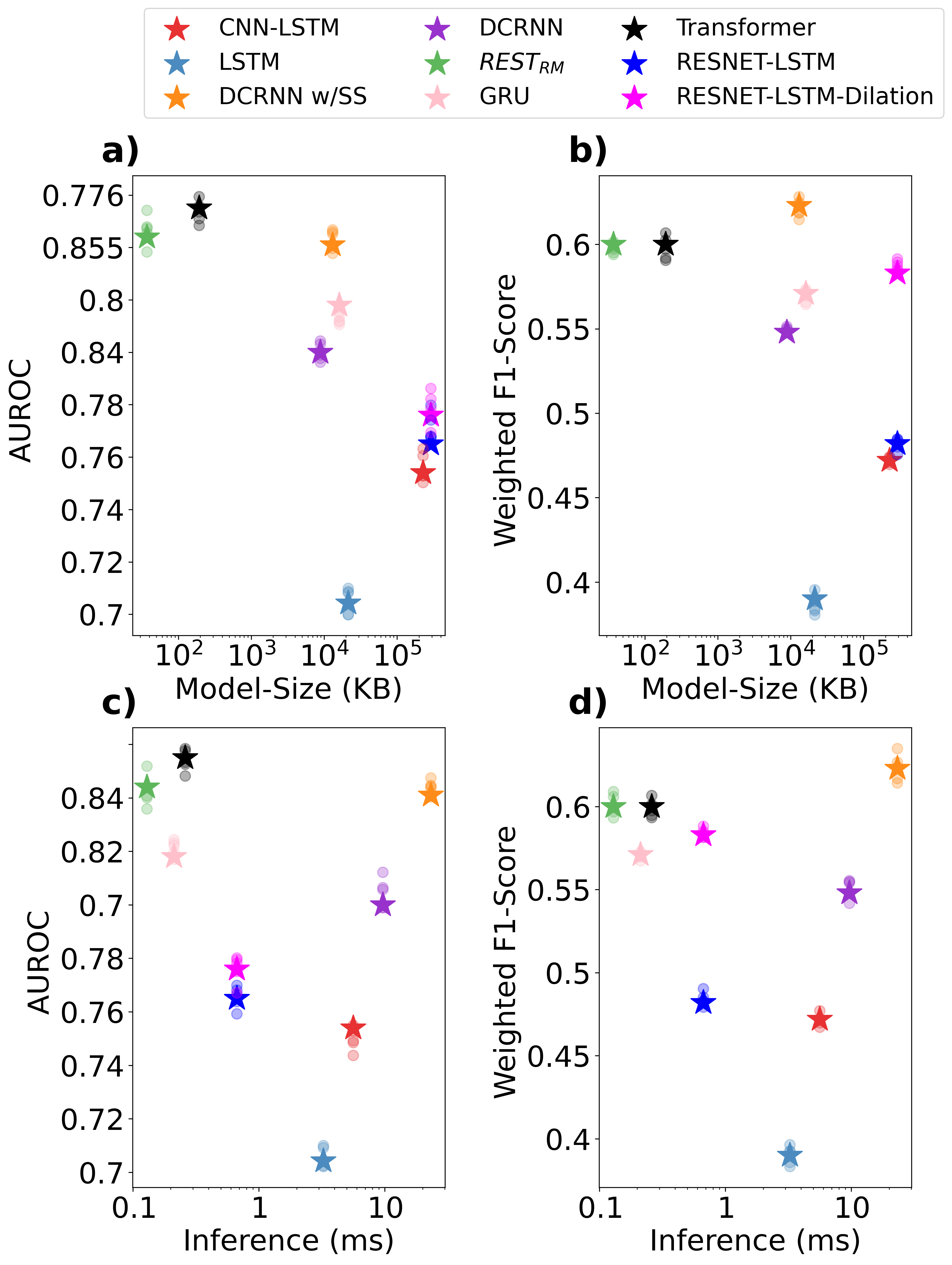}
  \caption{
Performance comparison in seizure analysis across models on TUSZ dataset: \textbf{a)} Seizure detection AUROC vs. Model size. \textbf{b)} Seizure classification weighted F1-score vs. model size. \textbf{c)} Seizure detection AUROC vs. inference. \textbf{d)} Seizure classification weighted F1-score vs. inference. The $\bullet$s represents the accuracy on evaluation set for different train/validation splits and  $\star$s represent the mean accuracy across different train/validation splits.}

  \label{fig:fig3}
  \end{minipage}
\end{figure}

\subsection{Experimental Results}

\textbf{Seizure Detection and Classification Accuracy:} We evaluated the performance of all baseline models and {\sc Rest} using the Area Under the Receiver Operating Characteristic Curve (AUROC) for seizure detection and Weighted F1-Score for seizure classification. Our model surpassed all baselines significantly on the CHB-MIT dataset for all different clip lengths. For the TUSZ dataset, it achieved very close detection AUROC scores for all clip lengths compared to DCRNN with self-supervision and the Transformer, while outperforming them at clip lengths of 6 and 8 seconds. \cref{fig:fig2} suggests that multiple random updates  improve the stability of {\sc Rest} as it leads to higher and more consistent performance compared to other models. According to \cref{fig:fig2}, {\sc Rest}\textsubscript{(RM)} and DCRNN with self supervision exhibit more stable performance over time across clip lengths, yielding consistent results. Interestingly, CNN-LSTM achieved higher performance in a small clip size of 4s, surpassing DCRNN with graph convolution layers.

\textbf{{\sc Rest} Enjoys an Exponentially Smaller Size:} 
While maintaining high accuracy, {\sc Rest} exhibits a size that is 14$\times$ smaller than the  smallest existing model for seizure detection and classification on \gls{tuhz} dataset. Table \ref{table:table3} highlights that {\sc Rest} requires 38$\times$
fewer parameters than state-of-the-art models (DCRNN w/SS) and over 697$\times$ fewer parameters than the deep CNN-LSTM model for seizure analysis.

Figure \ref{fig:fig3} a-b showcases {\sc Rest}'s outstanding performance, achieving an AUROC of 83.6\% for seizure detection with a clip length of 10 seconds. Additionally, {\sc Rest} secures the second-highest F1-Score for seizure classification, trailing only 2\% below DCRNN w/SS but with a significantly smaller size than all other baselines. The substantial gap between {\sc Rest}'s size and the sizes of other baselines, depicted on the logarithmic scale in Figure \ref{fig:fig3} a-b, underscores {\sc Rest}'s remarkable size advantage and potential for implementation on edge devices. The graph convolution layers in {\sc Rest} efficiently capture both short and long-range communication between nodes, ensuring high accuracy with a compact model size. Moreover, using identical weights for multiple random updates eliminates the need for additional layers while enhancing the model's accuracy and memory efficiency.

\textbf{Rapid Seizure Detection:}
{\sc Rest}\textsubscript{(RM)} achieves the fastest inference speed among all  models, being 20$\times$ faster than DCRNN w/SS and 9$\times$ faster than DCRNN during inference, with only a minor AUROC drop of less than 2\% for seizure detection across various clip lengths for \gls{tuhz} dataset. Moreover, {\sc Rest}, with multiple updates, requires only 1.292 ms for seizure detection, which is three times faster than the fastest baseline, \gls{lstm}, while being 13\% more accurate in delivering predictions (at 10-s clip length). On the CHB-MIT dataset, {\sc Rest} outperforms all other baselines in the seizure detection task, being the only model with an AUROC higher than 90\%. It also significantly outperforms other baselines for the short clip length of 4 seconds, which is crucial for real-time seizure detection \cite{zhu2021closed}.

In seizure classification, {\sc Rest}\textsubscript{(RM)} secures the second-highest F1-Score (Table \ref{table:table4}) and excels in providing the fastest classification result within 1.51 ms (Figure \ref{fig:fig3} c-d). Notably, it is three times faster than \gls{lstm}, while achieving 21\% higher accuracy than \gls{lstm}. 
The swift prediction capability of our model is attributed to its efficient design. {\sc Rest} relies on a single affine mapping into the state space, complemented by  two computationally lightweight graph convolutions. 

\section{Conclusion}
In this work, we propose {\sc Rest}, a graph-based residual state update mechanism for efficient seizure detection and classification tasks. Our model effectively captures both spatial and temporal behaviors of EEG signals, achieving  state-of-the-art performance in seizure detection and classification. With its shallow structure, {\sc Rest} boasts a fast inference speed, making it 9 times faster than current models with a comparable performance. Furthermore, {\sc Rest} exhibits remarkable efficiency, requiring only 37KB of memory, which is 14 times smaller than smallest existing models for seizure analysis tasks. These advancements position {\sc Rest} as a promising model for implementation on small, low-power edge devices, particularly for applications in epilepsy treatments like \gls{dbs} and \gls{rns}.

\section*{Impact Statement}
The EEG Seizure Corpus from Temple University Hospital, utilized in our research, is anonymized and publicly accessible with IRB approval \cite{tuh1,tuh2}. The authors declare no conflicts of interest, and the seizure detection and classification models presented in this study do not provide any harmful insights. Although our model has demonstrated accuracy in real-time seizure analyses, further experiments are essential for real-world application and implementation on edge devices, as demonstrated  in a number of recent systems \cite{shoaran2018energy, shin2022neuraltree, shaeri202433}. 
These evaluations should encompass testing with diverse datasets from various patient populations and hospitals. Additionally, assessing the model's energy efficiency is crucial to ensure its safety for chronic use, along with obtaining neurologists' approval regarding its neurological aspects for deployment in such devices.

\section*{Acknowledgements}
This work was supported in part by the Swiss State Secretariat for Education, Research and Innovation under Contract number SCR0548363, in part by the Wyss project under contract number 532932, in part by Hasler Foundation Program: Hasler Responsible AI project number 21043, in part by the Army Research Office under grant number W911NF-24-1-0048, and in part by the Swiss National Science Foundation (SNSF) under grant number 200021\_205011. Moreover, we appreciate the reviewers for their insightful feedback, which has significantly enhanced the robustness and clarity of our results.

\nocite{langley00}

\bibliography{refrences}
\bibliographystyle{icml2024}

\newpage
\appendix
\onecolumn

\section*{Appendix Introduction}

The Appendix is organised as followes:

\begin{itemize}
    \item Preprocessing details are outlined in \cref{ap:a}.
    \item The mathematical proof addressing the avoidance of gradient vanishing in our model is provided in \cref{ap:b}.
    \item Seizure analyses results are presented in \cref{ap:c}.
    \item Hyperparameter selection and training details for all models are discussed in \cref{ap:d}.
    \item The impact of BCE and MSE loss on training {\sc Rest} is compared in \cref{ap:e}.
    \item Training times are documented in \cref{ap:f}.
    \item Details explaining how {\sc Rest} avoids overfitting are shown in \cref{ap:g}.
    \item Differences between various graph structures are explored in \cref{ap:h}.
    \item Information about baseline compression is provided in \cref{ap:i}.
    \item F1-scores for seizure detection are presented in \cref{ap:j}.
    \item The effectiveness of binary random masking on different RNN variants is shown in \cref{ap:k}.
    \item Size comparisons for models with the same number of neurons are provided in \cref{ap:l}.
    \item Real-time evaluations of different models with overlapping windows are detailed in \cref{ap:m}.
    \item An ablation study on the inference performance of {\sc Rest} with and without binary random masking is presented in \cref{ap:n}.

\end{itemize}

\section{Details of Preprocessing}
\label{ap:a}
We initially performed general preprocessing on the EEG data followed by specific steps for each detection and classification tasks:

\subsection{TUSZ dataset}

\textbf{General Preprocessing:} The EEG signals in the TUH EEG Corpus (TUSZ) dataset were initially sampled at various frequencies. As a part of the preprocessing pipeline, all signals were uniformly resampled to 200 Hz. Subsequently, EEG clips were extracted using the natural choice of one-second, non-overlapping windows, resulting in an EEG tensor \(X \in \mathbb{R}^{T \times L \times N}\), where \(T\) represents clip lengths (ranging from 4, 6, 8, 10, 12, to 14 seconds), \(N\) is the number of electrodes (19), and \(L\) is the number of time samples (200). To harness the effectiveness of Fourier transform for neural EEG recordings, fast Fourier transform was applied to extract frequency components for each node at each time point. The log-amplitude of the frequencies was then computed and only non-negative frequency components were extracted  similar to prior studies \cite{tang2021self, ahmedt2020neural} leading to EEG clip tensor of \(X \in \mathbb{R}^{T \times M \times N}\) with $M$=100. Last, we have z-normalized the EEG clips across their second dimension for further analyses.  

\textbf{Preprocessing for Seizure Detection:} For seizure detection after extracting \gls{eeg} clips from the entire training set consisting of 5545 sessions, a binary label was assigned, with $y=1$ indicating the presence of at least one seizure within the clip and $y=0$ otherwise. To handle the issue of a substantial number of background clips in the dataset, non-seizure clips were randomly selected to achieve a balanced representation with seizure clips in the training data. Also, the last clip was dropped for each \gls{eeg} data if the recording ends before the clip could reach it's length.

\textbf{Preprocessing for Seizure Classification:} 
For seizure classification followed by \citet{tang2021self,ahmedt2020neural} we have removed the background data and only processed the seizure clips. We have started 2 seconds before the annotated seizure for tolerance in the annotations. Then we have labeled the clip $y=0$ for general non-specific (GN), $y=1$ for combined tonic (TC), $y=2$ for absence (AB), $y=3$ for focal, and $y=4$ for complex parietal (CP) seizures. Moreover, if  seizure event is shorter than the clip length we have truncated the clip to avoid having multiple seizures in one clip. Also, it is noteworthy that while the training set included simple partial seizures, these seizure types were absent in the evaluation set. Therefore, we excluded simple parietal seizures from the classification task. Also, because the clips for seizure classification may have different lengths we pad 0's to the end of the clip to assure all samples share the same length.

\subsection{CHB-MIT Dataset}
For the CHB-MIT dataset, we randomly selected 18 patients for training, 3 for evaluation, and 3 for testing. We followed the same preprocessing pipeline as described for the TUSZ dataset, with the exception of maintaining a uniform sampling rate of 256Hz for all patients. For each 1-second time window, we have 256 samples of raw EEG data per channel. The number of channels is consistent with the TUSZ dataset, comprising 19 channels, and we excluded any sessions with a different number of channels.

We utilized the same frequency domain components for seizure detection. Unlike the TUSZ dataset, the CHB-MIT dataset does not include seizure types for classification. The results are reported based on five different random seeds for the train/test/evaluation splits (more details at \cref{table:chbpat}).

\begin{table}[h]
\begin{center}

\caption{\label{table:chbpat} Details of sessions and number of seizures for each patient at CHB-MIT dataset.}
\vspace{0.2cm}
\begin{tabular}{l|ccc}

\textbf{Case} & {Number of Seizures} & {Number of Sessions} & {Age} \\
\Xhline{4\arrayrulewidth}
1 & 7 & 24 & 11 \\
\hline
2 & 3 & 36 & 11 \\
\hline
3 & 7 & 38 & 14 \\
\hline
4 & 4 & 42 & 22 \\
\hline
5 & 5 & 39 & 7 \\
\hline
6 & 10 & 18 & 1.5 \\
\hline
7 & 3 & 19 & 14.5 \\
\hline
8 & 5 & 20 & 3.5 \\
\hline
9 & 4 & 19 & 10 \\
\hline
10 & 7 & 25 & 3 \\
\hline
11 & 3 & 35 & 12 \\
\hline
12 & 27 & 24 & 2 \\
\hline
13 & 10 & 33 & 3 \\
\hline
14 & 8 & 26 & 9 \\
\hline
15 & 20 & 40 & 16 \\
\hline
16 & 8 & 19 & 7 \\
\hline
17 & 3 & 21 & 12 \\
\hline
18 & 6 & 36 & 18 \\
\hline
19 & 3 & 30 & 19 \\
\hline
20 & 8 & 29 & 6 \\
\hline
21 & 4 & 33 & 13 \\
\hline
22 & 3 & 31 & 9 \\
\hline
23 & 7 & 9 & 6 \\
\hline
24 & 16 & 22 & Unknown \\
\end{tabular}
\end{center}
\end{table}

\section{Preventing Gradient Vanishing with Residual Update}
\label{ap:b}
In equations \cref{eq:3,eq:4,eq:5}, the model's state is updated using a residual state update. When we take the derivative of $S^{t-1}$ concerning the forward propagation of \cref{eq:3}, we get:

\begin{equation}
\label{eq:13}
\frac{\partial\mathcal{L}}{\partial S^{t-1}} = \frac{\partial\mathcal{L}}{\partial S^{t}}\frac{\partial {S^t}}{\partial S^{t-1}} = \frac{\partial\mathcal{L}}{\partial S^{t}}\Big(1+\frac{\partial {\delta S^t}}{\partial S^{t-1}}\Big) = \frac{\partial\mathcal{L}}{\partial S^{t}} + \frac{\partial\mathcal{L}}{\partial S^{t}}\frac{\partial {\delta S^t}}{\partial S^{t-1}} .
\end{equation}

Here the $\mathcal{L}$ is the loss function to be minimized. This equation shows that the gradient of the previous state $S^{t-1}$ always has a term $\frac{\partial\mathcal{L}}{\partial S^{t}}$ directly added. This helps prevent the gradients of $\frac{\partial\mathcal{L}}{\partial S^{t-1}}$ from becoming too small, even when the gradients of the previous updates are small, i.e., $\frac{\partial\mathcal{L}}{\partial S^{t}}\frac{\partial {\delta S^t}}{\partial S^{t-1}}$.

\section{ROC Curves and Confusion Matrices for Different Clip Lengths}
\label{ap:c}
\begin{figure}[H]
\begin{minipage}[t]{1\columnwidth} 
 \includegraphics[width=\textwidth ]{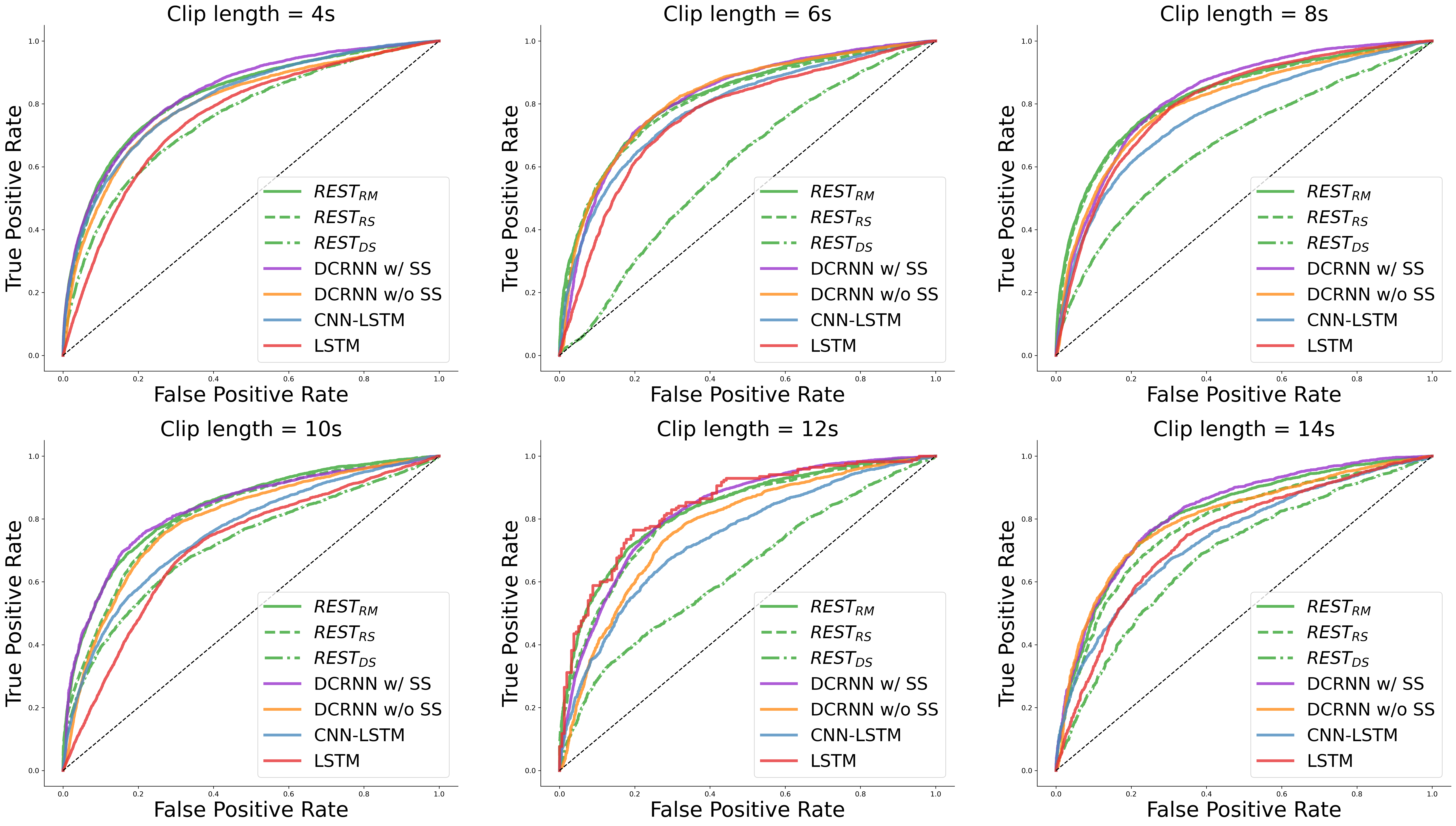}
  \caption{ ROC curves for different clip lengths among {\sc Rest} and baselines for TUSZ dataset.}
  \label{fig:fig4}
  \end{minipage}
\end{figure}

\begin{figure}[H]
\begin{minipage}[t]{1\columnwidth} 
 \includegraphics[width=\textwidth ]{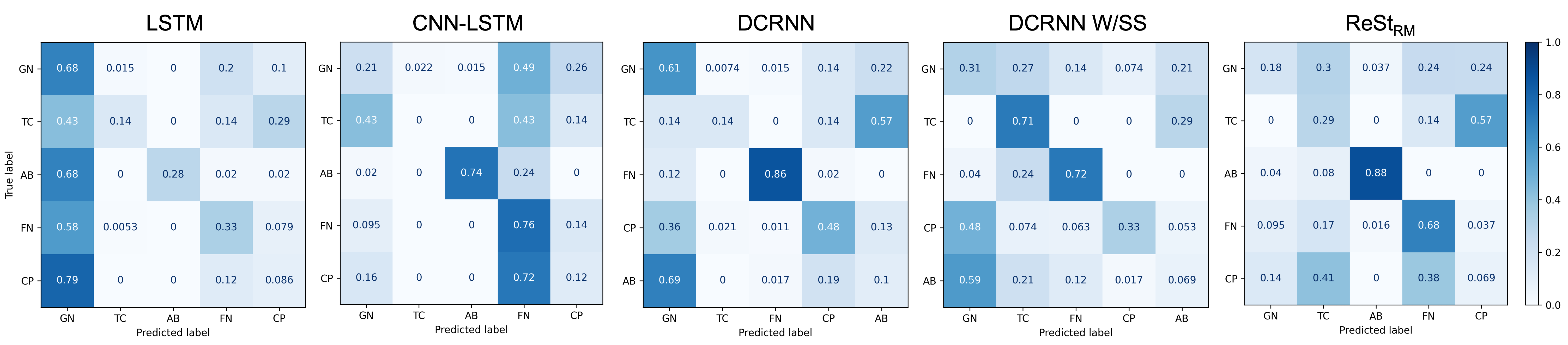}
  \caption{ Confusion Matrices for seizure classification task among different models.}
  \label{fig:fig5}
  \end{minipage}
\end{figure}

\section{Model Training and Hyperparameter Selection Details}
\label{ap:d}
Here are the details of training and hyperparameter selection for {\sc Rest} and baselines:

\textbf{{\sc Rest} Hyperparameters:}
We optimized the following hyperparameters for {\sc Rest} based on the lowest validation error:
a) Number of neurons in each graph convolution layer within the range [16, 32, 64];
b) Initial learning rate within the range [5e-4, 1e-4];
c) Success probability of the random binary mask within [0.1, 0.3, 0.5, 0.7, 1].
For multi-update {\sc Rest}, the number of updates for each time point was randomly selected an inteager from the interval [1, 10]. We conducted training for 500 epochs using a Multistep learning rate scheduler. Five experiments were run in PyTorch with different random seeds.

\textbf{DCRNN:} 
We followed the hyperparameter tuning strategy from the original paper \cite{tang2021self} for both DCRNN with and without self-supervision tasks. The hyperparameter search on the validation set included:
a) Initial learning rate within the range [5e-5, 1e-3];
b) Number of Diffusion Convolutional Gated Recurrent Units (DCGRU) layers within the range \{2, 3, 4, 5\} and hidden units within the range \{32, 64, 128\};
c) Maximum diffusion step K $\in$ \{2, 3, 4\};
d) Dropout probability in the last fully connected layer.
For self-supervised pre-training, we utilized mean absolute error (MAE) as the loss function. The models underwent training for 350 epochs with an initial learning rate of 5e-4, employing a maximum diffusion step of 1 and 64 hidden units in both the encoder and decoder. Moreover, cosine annealing learning rate scheduler \cite{coslr} was used as scheduler.

\textbf{CNN-LSTM:} For the baseline CNN-LSTM, we adopt the identical model architecture outlined in \citet{ahmedt2020neural}. This configuration employes two stacked convolutional layers with 32 kernels of size 3 $\times$ 3, one max-pooling layer of size 2 $\times$ 2, one fully-connected layer with an output neuron count of 512, two stacked \gls{lstm} layers with a hidden size of 128, and one additional fully connected layer.

\textbf{LSTM:} We employed two stacked \gls{rnn} layers, each with 64 hidden units, and an additional fully connected layer for the final prediction.

\textbf{GRU:} For the GRU model, we used same number of layers and hidden units as LSTM.

\textbf{ResNet-LSTM}:  We followed two versions with and without dilation described at \citet{lee2022real}.

\textbf{Transformer}: We implemented a two-layer multi-head attention mechanism with 64 embedding dimensions and 16 heads for the transformer architecture. Additionally, we utilized time positional encoding as introduced by \citet{atenneed} for the original positional encoding.

For detection task for all models binary cross entropy loss was used exept for {\sc Rest} which \gls{mse} performs slightly higher during the validation step. For classification task weighted binary cross entropy was employed due to the highly imbalancy among different seizure types.

\section{Comparison Between MSE and BCE loss for Training {\sc Rest}}
\label{ap:e}
{\sc Rest} was trained for seizure detection using both \gls{mse} and \gls{bce} loss functions. However, \gls{mse} outperformed \gls{bce} in terms of stability and accuracy. This advantage is attributed to \gls{bce}'s tendency for unbounded growth in classification logits, hindering residual updates and message passing between graph nodes, particularly in multi-update scenarios, as discussed in \citet{randazzo2020self}. As shown in Figure \ref{fig:fig4} \gls{mse} has less fluctuations and more stability in validation error during training compared to \gls{bce} loss when training {\sc Rest} with multiple updates.

\begin{table}[!h]
\begin{center}
    \caption{\label{table:table5} Detection Performance of {\sc Rest} with \gls{bce} and \gls{mse} loss functions for clip length of 10s on validation set.}
    \vspace{0.2cm}
\begin{tabular}{ p{3cm} c}
 \multicolumn{2}{c}{Seizure Detection performance } \\
 Loss Function & {AUROC }  \\
  \Xhline{4\arrayrulewidth}
{\sc Rest} BCE & 80.4 \\
{\sc Rest} MSE &  \textbf{83.6} \\
 \end{tabular}

\end{center}
\end{table}

\begin{figure}[H]
\begin{minipage}[t]{1\columnwidth} 
 \includegraphics[width=\textwidth ]{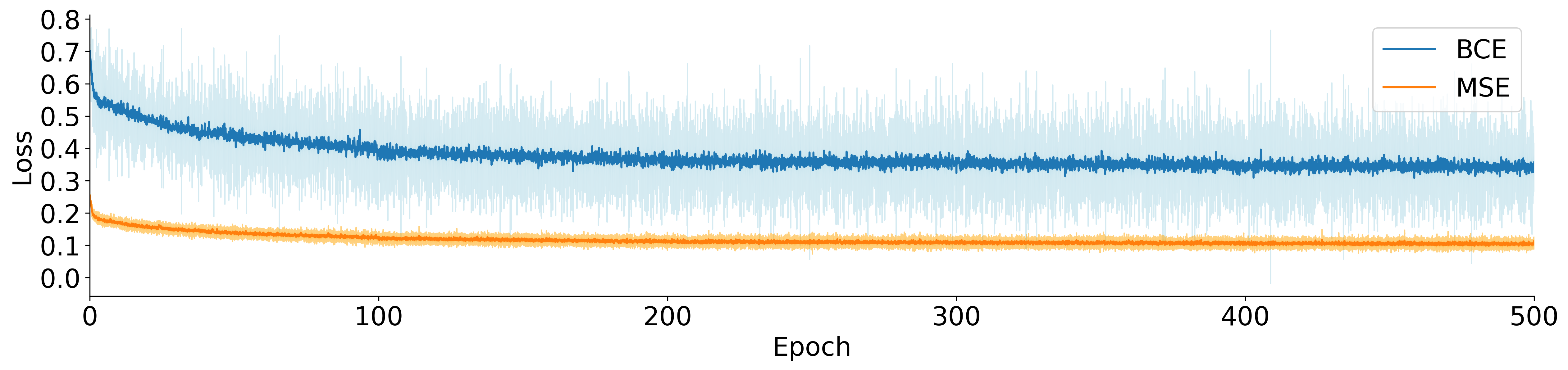}
  \caption{ Validation loss of the model for \gls{bce} and \gls{mse} loss functions.}
  \label{fig:fig6}
  \end{minipage}
\end{figure}

\section{Training Time}
\label{ap:f}

Bellow we report the time needed for training each model (\cref{table:table6}). All the models were trained on the same NVIDIA A100 GPU and the number of parameters and model size has reported at \cref{table:table4,table:table3}. {\sc Rest} requires more training time to adopt itself and converge to a stable point specially to adapt its update cell with multiple random updates

\begin{table}[h]
\begin{center}
\caption{\label{table:table6} Train time for seizure detection and classification tasks for different models among different clip lengths. Times are reported in minutes.}
\vspace{0.2cm}
\begin{tabular}{ p{4cm} p{.5cm} p{.5cm} p{.5cm} p{.7cm} p{.7cm} p{.7cm} || c}

 \multicolumn{7}{c}{Seizure Detection} & \multicolumn{1}{c}{Seizure Classification }\\

Model & 4-s & 6-s& 8-s & 10-s & 12-s & 14-s & 10-s\\
\Xhline{4\arrayrulewidth}

LSTM & 5 & 5 & 5 & 6 & 7 & 7 & 4\\
\hline
GRU & 5 & 5 & 5 & 6 & 7 & 8 & 4\\
\hline
CNN-LSTM & 8 & 8 & 8 & 9 & 9 & 10 & 5\\
\hline
ResNet-LSTM & 9 & 9 & 10 & 10 & 12 & 12 & 6\\
\hline
ResNet-LSTM-Dilation & 9 & 9 & 10 & 10 & 12 & 12 & 6\\
\hline
DCRNN & 20 & 22 & 23 & 25 & 28 &  30 & 20\\
\hline
DCRNN w/SS & 23 & 30 & 35 & 40 & 48 &  60 & 35\\
\hline
Transformer & 12 & 12 & 13 & 14 & 14 & 16 & 8\\
\hline
{\sc Rest}\textsubscript{(DS)} & 45 & 47 & 50 &  53 & 55 & 60 & 10\\
{\sc Rest}\textsubscript{(RS)}  & 45 & 47 & 50 &  53 & 55 & 60 & 10 \\
{\sc Rest}\textsubscript{(RM)}  & 70 & 75 & 80 &  90 & 95 & 100 & 25\\

 \end{tabular}
\end{center}
\end{table}

\section{{\sc Rest} Combat Forgetting at Each Time Point }
\label{ap:g}
While updating {\sc Rest} specially when the update cell includes multiple updates {\sc Rest} avoids forgetting the input by updating its state based on the affine mapping of the previous state and the input. As an example we consider two following setting: 

\textbf{Setting 1}: Updating the state based on previous state only where first the state is initialized as $S^t_i = WX^t + US^{t}_i $ and then it will iteratively update the state $S^t_i$ as follows:

\begin{equation}
\label{eq:14}
\delta S^{t}_i =  \mathcal{G}_\Theta(S^t_i),
\end{equation}
\begin{equation}
\label{eq:15}
    S^{t}_{i+1} = S^{t}_i + \delta S^{t}_{i}\odot B.
\end{equation}
\textbf{Setting 2:} Updating the state based on affine mapping of current input and previous state for iteratively update the state $S^t_i$ as follows:
\begin{equation}
\label{eq:16}
    H^t_i = WX^t + US^{t}_i 
\end{equation}
\begin{equation}
\label{eq:17}
\delta S^{t}_i =  \mathcal{G}_\Theta(H^t_i),
\end{equation}
\begin{equation}
\label{eq:18}
    S^{t}_{i+1} = H^{t}_i + \delta S^{t}_{i}\odot B.
\end{equation}

In Setting 1, after mapping from the input to the state space, the state is updated only based on the previous state. This setup poses a risk of the model forgetting information from the current input, especially if the update cell iteratively modifies the state multiple times. This situation hinders state from converging to a stable point and simply diverges due to neglecting the input data. In Setting 2, represented by {\sc Rest}'s update cell, the input plays a crucial role and is actively involved in the iterative update process, as shown in equations \cref{eq:16,eq:17,eq:18}. This design prevents the model from forgetting information from the current time input $X^t$, promoting convergence of the state to a more meaningful final state by utilizing the input's information throughout the updates.

\begin{figure}[H]
\begin{minipage}[t]{1\columnwidth} 
 \includegraphics[width=\textwidth ]{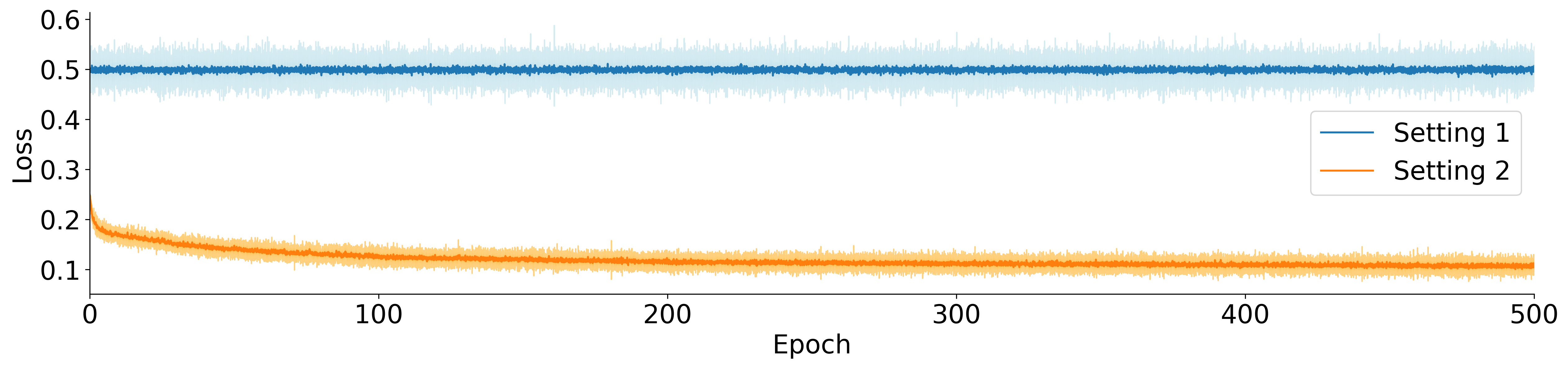}
  \caption{ Validation loss of setting 1 and setting 2 mentioned in the \cref{ap:f}.}
  \label{fig:fig7}
  \end{minipage}
\end{figure}

As illustrated in \cref{fig:fig7}, Setting 1 fails to converge to a stable point, and the validation loss remains unchanged throughout the training process.

\section{Comparison Between Different Gaussian Kernels Threshold for EEG Distance Graph}
\label{ap:h}
Here we illustrate different distance graph constructions based on different thresholds or the Gaussian kernel. The lower $k$ values (i.e. 0.6) results in missing connection between nodes and large $k$ thresholds results in connecting nodes which are far away. Similar to \citet{tang2021self} we also choose $k=0.9$ as threshold which resembles the \gls{eeg} montage (longitudinal bipolar and transverse bipolar) \cite{montage} and results in a reasonable node connection.

\begin{figure}[H]
\begin{minipage}[t]{1\columnwidth} 
 \includegraphics[width=\textwidth ]{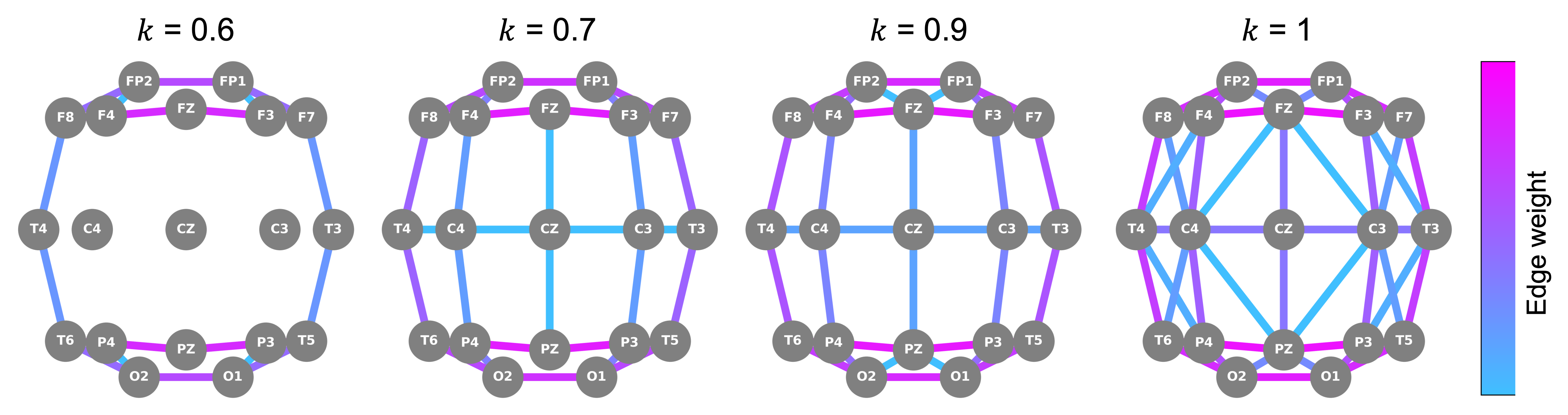}
  \caption{Illustration of distance based EEG graph constructed from different thresholds for the Gaussian kernel. $k$=0.9 was chosen for this study.}
  \label{fig:fig8}
  \end{minipage}
\end{figure}

\section{Compressing Baseline Models}
\label{ap:i}

We tried to compress existing models for seizure detection and classification, achieving performance comparable to those described in \citet{tang2021self}. However, in case of \gls{lstm} and CNN-LSTM models shrinking the model size without a significant performance drop proved challenging. We matched the performance reported in \citet{tang2021self} for DCRNN and DCRNN w/SS models with only one diffusion convolution gated recurrent unit, and reduced the model size by half, from 2.7 MB to less than 1 MB for DCRNN. Furthermore, for the seizure detection task, we achieved the same accuracy with 126K parameters compared to the original paper's 168,641 parameters.

For the classification task, the original paper \cite{tang2021self} reported 280,964 parameters for DCRNN and 417,572 parameters for DCRNN w/SS. In our compressed models, we achieved 126K parameters for DCRNN and 330K for DCRNN w/SS, successfully reducing the model size by a factor of 2 for DCRNN and a factor of 1.5 for DCRNN w/SS.

Despite successful reductions, the compressed models still possess a considerable number of parameters, especially in the presence of a gating mechanism, highlighting the non-parameter efficiency and memory demands associated with existing models for seizure detection.

\section{F1-Score for Seizure detection}
\label{ap:j}
Below is the F1-score (weighted averaged) results for seizure detection task on TUSZ dataset.

\begin{table}[ht]
\centering
\caption{F1-Score for seizure detection across different time windows for various models.}
\vspace{0.2cm}
\begin{tabular}{l|cccccc}
\textbf{Model} & {4-s} & {6-s} & {8-s} & {10-s} & {12-s} & 14-s \\
\Xhline{4\arrayrulewidth}
LSTM & 82.3 & 69.9 & 79.5 & 80.5 & 72.7 & 73.2 \\ 
\hline
CNN-LSTM & 70.1 & 69.5 & 75.3 & 73.5 & 68.3 & 67.5 \\ 
\hline
GRU & \textbf{82.7} & 69.9 & 81.6 & 80.5 & \textbf{81.0} & 71.3 \\ 
\hline
RestNet-LSTM & 79.7 & 78.2 & 80.1 & 75.1 & 77.0 & 76.3 \\ 
\hline
RestNet-Dilation-LSTM & 80.5 & 80.4 & 79.0 & 76.6 & 75.0 & 74.6 \\ 
\hline
Transformer & 78.45 & 79.3 & 78.5 & \textbf{82.0} & 79.1 & \textbf{79.2} \\ 
\hline
DCRNN & 81.2 & 80.2 & 81.6 & 80.0 & 74.2 & 72.0 \\ 
\hline
DCRNN W/SS & 75.2 & \textbf{81.1} & 81.2 & 81.0 & 75.7 & 76.0 \\ 
\hline
Rest(RS) & 69.5 & 68.4 & 78.3 & 79.1 & 74.7 & 74.1 \\ 
\hline
Rest(RM) & 81.0 & 75.2 & \textbf{83.2} & 81.0 & 75.7 & 76.2 \\ 
\end{tabular}

\label{tab:performance_metrics}
\end{table}

\section{Binary Random Masking and Multiple Updates for Other RNNs}
\label{ap:k}
We conducted an ablation study to evaluate the performance of RNN baselines with single and multiple random updates, as shown in Table \ref{tab:vanilla_rs_rm}.

\begin{table}[h]
\centering
\caption{AUROC for seizure detection on a window size of 10s for the TUSZ dataset. Vanilla models are RNN variants without any update techniques, while RM and RS are {\sc Rest} update strategies.}
\vspace{0.2cm}

\begin{tabular}{l|ccc}
\textbf{Model} & {Vanilla} & {RS} & {RM} \\ 
\Xhline{4\arrayrulewidth}
RNN & 77.3 & 80.1 & \textbf{80.8} \\ 
\hline
GRU & 73.5 & 72.8 &\textbf{ 73.6} \\ 
\hline
LSTM & 70.4 & 74.5 & \textbf{74.7 }\\ 
\end{tabular}

\label{tab:vanilla_rs_rm}
\end{table}

As shown, the RNN variants can improve their performance in seizure detection tasks using {\sc Rest} update techniques.

\section{Size Comparison with 64 Number of Neurons for all Models}
\label{ap:l}

\begin{table}[ht]
\centering

\caption{Size comparison of different models using an equal number of neurons (64). The table indicates the model size and parameter count for {\sc Rest} and baseline models.}

\vspace{0.2cm}

\begin{tabular}{l|cc}
\textbf{Model} & {Parameters (\#)} & {Size (MB)} \\ 
\Xhline{4\arrayrulewidth}
DCRNN w/SS & 330K & 1.319 \\ 
\hline
DCRNN  & 126K & 0.884 \\ 
\hline
Transformer & 48.3K & 0.193 \\ 
\hline
GRU & 402K & 1.61 \\ 
\hline
ResNet-LSTM & 7.5M & 30.3\\ 
\hline
ResNet-LSTM-Dilation  & 7.5M & 30.3 \\ 
\hline
LSTM & 536K & 2.147 \\ 
\hline
CNN-LSTM & 6M & 22.8 \\ 
\hline
REST(DS) & 27K & 0.051 \\ 
\hline
REST(RS) & 27K & 0.051 \\ 
\hline
REST(RM) & 27K & 0.051 \\ 

\end{tabular}

\label{tab:model_parameters_sizes}
\end{table}

\section{More Evaluation for Real-Time Detection}
\label{ap:m}

We followed the real-time seizure detection framework described by \citet{lee2022real}, using a 4-second clip length for seizure detection with a 3-second overlap between consecutive clips. We measured both the inference time and latency, the latter being the delay between the actual onset of a seizure and the model's detection. Low latency is crucial to avoid late detection of seizure events. As shown in \cref{tab:lat}, {\sc Rest} achieves the lowest latency alongside the Transformer model, while also maintaining significantly lower inference times compared to all other baselines.

\begin{table}[h]
\centering
\caption{Comparison of different models' performance on real-time seizure detection with overlapping windows}

\vspace{0.3cm}
\label{tab:lat}
\begin{tabular}{l|ccc}
\textbf{Model} & {AUCROC} &  Latency (s)& {Inference (ms)} \\
\Xhline{4\arrayrulewidth}
LSTM & 75.5 & 0.31 & 3.254 \\
\hline
GRU & 76.1 & 0.4 & 2.12 \\
\hline
RestNet-LSTM & 79.1 & 0.3 & 6.78 \\
\hline
RestNet-Dilation-LSTM & 80.2 & 0.34 & 6.78 \\
\hline
CNN-LSTM & 81.3 & 0.26 & 5.624 \\
\hline
DCRNN & 79.7 & 0.25 & 9.67 \\
\hline
Transformer & \textbf{83} & \textbf{0.2} & 2.5 \\
\hline
\sc{Rest}(DS) & 75.3 & 0.23 & \textbf{0.615} \\
\hline
\sc{Rest}(RS) & 79.4 & \textbf{0.2} & 0.71 \\
\hline
\sc{Rest}(RM) & 82.4 & 0.25 & 1.29 \\

\end{tabular}
\end{table}

\section{{\sc Rest} W/O Binary Random Mask during Inference}
\label{ap:n}

We evaluated {\sc Rest} performance with and without masking over the inference in which similar to \citet{dropout} strategy the mask was removed and the incremental state update was scaled using the success probability of the binary mask ($p$).

\begin{table}[ht]
\centering
\caption{Performance metrics for seizure detection on 10-s clip length with and without inference mask}
\vspace{0.2cm}

\begin{tabular}{l|cc}
 
\textbf{Model} & {W/ Inference Mask} & {W/O Inference Mask} \\ 
\Xhline{4\arrayrulewidth}
{\sc Rest} (RS) & \textbf{81.8} & 81.5 \\ 
\hline
{\sc Rest} (RM)  & \textbf{83.6} & 82.9 \\ 

\end{tabular}
\label{tab:rest_inference_mask}
\end{table}


\end{document}